\documentclass[journal]{IEEEtran}

\usepackage{myStyleIEEE}
\usepackage{nomencl}
\usepackage{ragged2e}
\IEEEoverridecommandlockouts

\newlength{\bracewidth}

\theoremstyle{definition}
\newtheorem{definition}{Definition}

\usepackage{etoolbox}
\renewcommand\nomgroup[1]{%
  \item[\bfseries
  \ifstrequal{#1}{P}{A. Parameters}{%
  \ifstrequal{#1}{V}{C. Variables}{%
  \ifstrequal{#1}{S}{B. Sets and Indices}{}}}%
]}

\begin{document}

\title{Headroom as A Grid Service in Software-Defined Power Grids: A Peak-to-Peak Control Design Approach}

\newtheorem{proposition}{Proposition}
\renewcommand{\theenumi}{\alph{enumi}}

\author{Zhongda~Chu,~\IEEEmembership{Member,~IEEE,} and
        Fei~Teng,~\IEEEmembership{Senior Member,~IEEE} 
        
        
\vspace{-0.5cm}}
\maketitle
\IEEEpeerreviewmaketitle

\begin{abstract}
To address system frequency challenges driven by the integration of renewable generation, advanced control strategies are designed at the device level to provide effective frequency support following disturbances. However, typically relying on energy-based performance metrics, these methods cannot guarantee the system frequency constraints such as frequency nadir and maximum Rate-of-Change-of-Frequency (RoCoF). Moreover, locally-designed frequency support cannot minimize the overall system cost to maintain frequency stability. On the other hand, the concept of frequency-constrained system scheduling is introduced, which incorporates frequency dynamic constraints into the system economic optimization, so that frequency requirements can be maintained with minimum cost. However, these works rely on analytical approximations of the frequency dynamic metrics, which are mathematically complicated and tend to be over-conservative for the approximation of IBR headroom requirements. 
This paper resolves these challenges by proposing a peak-to-peak control design, which not only confines the frequency nadir and maximum RoCoF, but also significantly reduces the complexity of the additional frequency constraints by converting the determination of optimal IBR control parameters to that of the IBR headroom reserve, leading towards a new grid service. The overall performance of the proposed method is demonstrated in the modified IEEE-39 bus system. 
\end{abstract}


\makenomenclature
\renewcommand{\nomname}{List of Symbols}
\mbox{}
\nomenclature[P]{$M_g$}{system inertia$\,[\mathrm{MWs/Hz}]$}

\section{Introduction} \label{sec:1}
The rapid integration of renewable energy sources into modern power systems has brought about significant challenges \cite{milano2018foundations}. Traditional power system frequency regulation has relied heavily on synchronous generators, which inherently provide inertia and frequency support. However, due to the power electronic interface, IBRs lack this inherent capability, thus leading to severe frequency deviations and Rate-of-Change-of-Frequency (RoCoF). This can cause undesired events such as under-frequency load shedding, cascade failure, and compromise system reliability. 

To address these challenges, grid-forming control strategies, such as the Virtual Synchronous Machine (VSM), have been developed \cite{muftau2022role}. These strategies enable Inverter-Based Resources (IBRs) to emulate the inertia and frequency response characteristics of traditional synchronous machines, thereby providing essential frequency support to the grid. The control parameters of the VSMs (virtual inertia and damping) are typically determined by offline optimal control design algorithms. Specifically, the authors in \cite{xie2024game} utilize a novel distributed game-theoretic learning approach to coordinate aggregators in the power system for fast frequency response provision. A fractional-order model prediction controller is presented in \cite{long2023frequency} for fractional-order VSMs to alleviate the output power oscillation and achieve an optimal frequency and voltage regulation for the grid. An LQR-based adaptive controller is designed to determine the optimal VSM parameters in \cite{markovic2018lqr}, while considering the different needs at various stages during the frequency event. An $\mathcal{H}_2$ norm-based optimization problem is formulated in \cite{poolla2019placement} to optimize the parameters and locations of VSMs in a power system such that the system frequency deviation and RoCoF can be minimized. The above-mentioned control design utilizes energy-based performance metrics of the system states (frequency deviation and RoCoF) as well as the control input to balance the trade-off between the performance and control efforts. However, there lacks a connection between these energy-based metrics and the maintenance of the system-level frequency constraints, e.g., the frequency nadir and maximum RoCoF. Moreover, during a severe frequency event, priority is given to maintaining the frequency deviation and RoCoF within the permissible bounds, whereas the energy consumption due to the control input is of less significance. Additionally, most of the VSM controls are designed locally on the device level and the system-level optimality cannot be achieved without the coordination among different resources.

On the other hand, the concept of frequency-constrained optimization has also been proposed to minimize the overall system operation/investment cost while ensuring system-level frequency stability. 
The authors in \cite{7115982} propose a full stochastic scheduling model that simultaneously optimizes energy production as well as the frequency services to ensure the system frequency constraints while considering the uncertainty of the renewable generation. 
In \cite{yang2022distributionally}, a scheduling framework for integrated electricity-gas systems is presented, which accounts for frequency constraints alongside operational limitations of the gas network. To address wind power uncertainty, the model employs distributionally robust joint chance constraints. Meanwhile, \cite{zhang2022frequency} develops a frequency security-constrained scheduling method that incorporates the frequency support and reserve provision from wind farms. The support capability is precisely characterized by accounting for the actual grid-connected wind turbine capacity and the influence of wake effects.


Although these works consider the frequency support from IBRs in low-inertia systems, they assume the frequency support is provided through a pre-defined control structure and control parameters, without fully capitalizing the flexibility offered by the software-defined nature of IBRs.
As a result, the optimality of the overall system cannot be ensured considering the time-varying operating conditions of the system. The problem can be solved by viewing the control parameters of the IBR frequency support as decision variables in the system scheduling model. The authors in \cite{9066910} introduce a frequency-constrained scheduling model where the optimal frequency control parameter of the IBRs can be optimally determined to ensure the system frequency stability, while considering the IBR control capability. Similarly, \cite{shen2023optimal} proposes a frequency-constrained stochastic look-ahead power dispatch model to formulate the frequency control parameters of IBRs as scheduling variables, which can optimally allocate the virtual inertia and droop coefficient in the system. However, the feasible ranges of the control parameters are either oversimplified or assumed to be known. 

In this context, this paper incorporates the peak-to-peak control design into the frequency-constrained scheduling model, which not only converts the conventional highly nonlinear frequency nadir constraints to a simple linear one but also changes the decisions from the virtual inertia and damping provided by IBRs to the IBR headroom reserve, as a new grid service. This fits the existing framework of day-ahead ancillary service market while allowing the control parameters of IBRs to be redesigned in the real-time dispatch stage, fully utilizing the software-defined nature of IBRs.
The main contributions of this work are identified as follows.
\begin{itemize}
    \item A novel peak-to-peak control design is proposed for the frequency regulation in IBR-dominated power systems, where the conservativeness, in terms of the distance between the system state trajectory and the state-invariant ellipsoid, is significantly reduced through the change of coordination.
    \item The IBR headroom reserve required to support the system frequency dynamics is identified by solving semidefinite programming (SDP) iteratively, which can be defined as a grid service,  while the exact control parameters can be revealed afterwards according to system needs closer to real time, benefiting from the software defined nature of IBRs. 
    \item A novel frequency-constrained system scheduling model is proposed where the highly nonlinear frequency nadir constraint is replaced by that of the IBR headroom reserve requirement, which significantly reduces the complexity of the optimization model.
    
    \item The effectiveness of the proposed model is verified in the modified IEEE 39-bus system with the benefit in system operation cost, computational efforts and the uncertainty management being demonstrated.
\end{itemize}
The rest of the paper is organized as follows. Section~\ref{sec:2} introduces the frequency dynamics in high IBR-penetrated systems and peak-to-peak control design to determine the optimal IBR control parameters. Section~\ref{sec:3} presents the system scheduling model, with the frequency and IBR reserve constraints being developed. Case studies are discussed in Section~\ref{sec:4} followed by conclusions in Section~\ref{sec:5}.

\section{System Frequency Dynamics} \label{sec:2}
This section begins by presenting the analytical formulation of frequency dynamics in a general power system with high penetration of IBRs. Subsequently, a peak-to-peak control strategy is employed for frequency regulation to determine the optimal virtual inertia and damping from IBRs, along with the corresponding headroom reserve requirements.

\subsection{Frequency Dynamics in Low Inertia Systems} \label{subsec:2.1}
\begin{figure}[!b] 
	\centering
	\scalebox{0.9}{\includegraphics[]{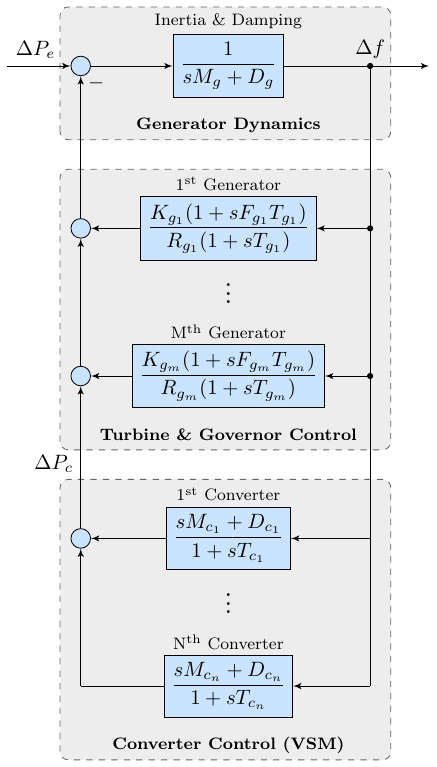}}
	\caption{Frequency dynamics of general multi-machine systems.}
	\label{fig:freq_dyn}
\end{figure}

Consider a power system comprising both conventional (denoted by subscript $g$) and converter-based (denoted by subscript $c$) generators, as illustrated in Fig.~\ref{fig:freq_dyn}. The dynamic behavior of the generators is governed by the swing equation, where $M_g$ and $D_g$ represent the normalized inertia and damping coefficients of the conventional generators, respectively.
\begin{equation}
    M_g=\dfrac{\sum_{i\in\mathcal{N}_g}M_{g_i}P_{g_i}}{P_b} \quad , \quad D_g=\dfrac{\sum_{i\in\mathcal{N}_g}D_{g_i}P_{g_i}}{P_b}.
\end{equation}
Let $i \in \mathcal{N}_g$ denotes the set of traditional synchronous generators, with $P_{g_i}$ representing their nominal power ratings. The base power, $P_b$, is defined as the total nominal capacity of all online generators. The turbine and governor control of SGs are modeled using the low-order model proposed in \cite{Anderson1990,Denis2006}, with the detailed parameter definitions in \cite{9066910}.
In addition, IBRs are incorporated to provide frequency support via a grid-forming VSM control scheme. Note that only the outer control loop associated with the active power is considered in the system frequency model, which is a common practice for frequency stability studies.  Let $\mathcal{N}_v$ denote the set of converters implementing VSM control. As illustrated in Fig.~\ref{fig:freq_dyn}, $T_{c_j}$, $M_{c_j}$ and $D_{c_j}$ are the time constants, virtual inertia and damping of each converter $j \in \mathcal{N}_v$, respectively.

\subsection{Frequency Nadir Derivation} \label{subsec:2.2}
Based on the system representation in Fig.~\ref{fig:freq_dyn}, the transfer function $G(s)$ characterizing frequency dynamics can be derived as:

\begin{align}
    G(s) &= \dfrac{\Delta f}{\Delta P_e} = \left(\underbrace{(sM_g+D_g)+\sum\limits_{i\in\mathcal{N}_g}\dfrac{K_{g_i}(1+sF_{g_i}T_{g_i})}{R_{g_i}(1+sT_{g_i})}}_{\text{synchronous generators}} \right. \nonumber \\
    &+ \left. \underbrace{\sum\limits_{k\in\mathcal{N}_v}\dfrac{sM_{c_k}+D_{c_k}}{1+sT_{c_k}}}_{\text{VSM converters}}\right)^{-1} \label{eq:G1}
\end{align}
It is reasonable to assume identical time constants for all synchronous generators, i.e., $T_{g_i} = T$ and $T \gg T_{c_{j,k}} \approx 0$ \cite{Ahmadi2014,9066910}. Under these assumptions, the transfer function in \eqref{eq:G1} can be simplified as follows:

\begin{equation}
    G(s) = \frac{1}{MT}\frac{1+sT}{s^2+2\zeta\omega_n s + \omega_n^2}. \label{eq:G2}
\end{equation}
The natural frequency $(\omega_n)$ and the damping ratio $(\zeta)$ are calculated as follows:
\begin{equation}
    \omega_n = \sqrt{\frac{D+R_g}{MT}} \quad , \quad \zeta = \frac{M+T(D+F_g)}{2\sqrt{MT(D+R_g)}} \label{eq:wn}
\end{equation}
where the related parameters take the form below:
\begin{subequations}
\begin{align}
    M & = M_g + M_c \label{eq:inertia}\\
    D & = D_g + D_c\label{eq:damping}\\
    F_g &= \sum\limits_{i\in\mathcal{N}_g}\dfrac{K_{g_i}F_{g_i}}{R_{g_i}}\dfrac{P_{g_i}}{P_b} \label{eq:term1}\\
    R_g &= \sum\limits_{i\in\mathcal{N}_g}\dfrac{K_{g_i}}{R_{g_i}}\dfrac{P_{g_i}}{P_b} \\
    M_c &= \sum\limits_{k\in\mathcal{N}_v}M_{c_k}\dfrac{P_{c_k}}{P_b} \\
    D_c &= \sum\limits_{k\in\mathcal{N}_v}D_{c_k}\dfrac{P_{c_k}}{P_b} .\label{eq:term5}
\end{align}
\end{subequations}
The terms $P_i/P_b$ in \eqref{eq:term1}–\eqref{eq:term5} result from expressing quantities in the per-unit system. As shown in \eqref{eq:inertia} and \eqref{eq:damping}, both the virtual inertia and damping introduced via VSM control, collectively influence the system’s total inertia and damping characteristics.



Considering a step disturbance in active power, represented as $\Delta P_e(s) = \Delta P_L/s$, the corresponding solution in for the frequency deviation $(\omega(t) \equiv \Delta f(t))$ is obtained as follows:
\begin{align}
    \omega(t) &= \frac{\Delta P_L}{MT\omega_n^2} \label{eq:wt} \\
    &+\frac{\Delta P_L}{M\omega_d}e^{-\zeta\omega_n t} \left( \sin{(\omega_d t) - \frac{1}{\omega_n T}\sin{(\omega_d t + \phi)}} \right) ,\nonumber
\end{align}
where $\omega_d$ and $\phi $ are defined as:
\begin{equation}
    \omega_d = \omega_n\sqrt{1-\zeta^2} \quad , \quad \phi = \text{sin}^{-1}\left(\sqrt{1-\zeta^2}\right). \label{eq:wd_phi}
\end{equation}
The time at which the frequency nadir occurs, denoted as $(t_m)$, can be identified by locating the point where the derivative of the frequency (RoCoF) equals zero:

\begin{align}
    \dot{\omega}(t_m) = 0 \longmapsto t_m = \frac{1}{\omega_d}\text{tan}^{-1}\left( \frac{\omega_d}{\zeta\omega_n - T^{-1}} \right)  \label{eq:tm}
\end{align}
Setting $t=t_m$ in \eqref{eq:wt} gives the following frequency nadir expression:


\begin{equation}
    \omega_\text{max} = \frac{\Delta P_L}{D+R_g} \left( 1 + \sqrt{\dfrac{T(R_g-F_g)}{M}} e^{-\zeta\omega_n t_m} \right) \label{eq:nadir}
\end{equation}
RoCoF reaches its maximum at $t_r=0^+$:
\begin{equation}
    \dot{\omega}_\text{max} = \dot{\omega}(t_r) = \frac{\Delta P_L}{M} .\label{eq:RoCoF}
\end{equation}
Examination of \eqref{eq:nadir}–\eqref{eq:RoCoF} reveals that the frequency nadir and RoCoF are influenced by the system’s total inertia and damping. These parameters can be regulated by the frequency support from IBRs. Specifically, RoCoF exhibits an inverse relationship with inertia, i.e., $\dot{\omega}_{\text{max}}\sim M^{-1}$, while the frequency nadir is governed by a nonlinear function of both inertia and damping, expressed as $\omega_{\text{max}}=f_{\omega}\left(M,D\right)$ in \eqref{eq:nadir}.


\subsection{Peak-to-Peak Control Design}
We first derive the state-space model of the concerned system. After the disturbance has been applied, i.e., $\forall t\in [0^{+},+\infty)$ $\Delta P_L$ can be considered a constant, thus transforming \eqref{eq:G2} into the following expression:
\begin{align}
    \ddot{\omega} = -2\xi\omega_n \dot{\omega} - \omega_n^2 \omega + \frac{\Delta P_L}{MT} \label{eq:w_rocof}.
\end{align}
Different from the conventional SGs, the frequency support from IBRs ($\Delta P_c$) depends on their control parameters ($M_c$ and $D_c$), which can be updated frequently according to the system operating conditions to achieve optimal performance. Combining \eqref{eq:wn} with \eqref{eq:w_rocof} and viewing $\Delta P_c$ as control input yield a state-space representation of the form:
\begin{subequations}
\label{eq:ss_general}
\begin{align}
\begin{bmatrix}
    \dot{\omega} \\ 
    \ddot{\omega}
\end{bmatrix} \nonumber
& = 
\underbrace{\begin{bmatrix}
    0 & 1 \\
    -\frac{D_g+R_g}{MT} & -\frac{M_g+T(D+F_g)}{MT}
\end{bmatrix}}_{A}
\begin{bmatrix}
    \omega \\ 
    \dot{\omega}
\end{bmatrix}
+
\underbrace{\begin{bmatrix}
    0 \\ \frac{1}{T M}
\end{bmatrix}}_{B_1} \Delta P_c \\
& +
\underbrace{\begin{bmatrix}
    0 \\ \frac{1}{T M}
\end{bmatrix}}_E \Delta P_L \label{eq:ss_g1} \\
\Delta P_c & = \underbrace{
\begin{bmatrix}
    -D_c & -M_c
\end{bmatrix}}_{K}
\begin{bmatrix}
    \omega \\ \dot \omega
\end{bmatrix} \label{eq:ss_g2}
\end{align}    
\end{subequations}
with $ x= \begin{bmatrix} \omega & \dot{\omega} \end{bmatrix}^T$ and $x(0) = \begin{bmatrix}0 &\dot\omega(0^+)\end{bmatrix}^T$ being the state vector, initial condition respectively. Defining $\eta = \Delta P_L$ and $u = \Delta P_c$ as the disturbance and the control input respectively enables us to further denote the system dynamics in a general form together with the output $z$:
\begin{subequations}
\label{eq:ss_open}
    \begin{align}
        \dot x &= Ax+B_1 u + E\eta \label{eq:ss-dx} \\
        z &= Cx + B_2 u \\
        u &= Kx,
    \end{align}
\end{subequations}
where $C$ and $B_2$ are the performance matrices with comfortable dimensions. However, due to the definition in \eqref{eq:inertia} and \eqref{eq:damping}, the matrix $A$, $B_1$ and $E$ also contain $M_c$ and $D_c$, hence being dependent on the feedback control gain $K$. An iterative approach is further proposed to solve this issue as discussed in Section~\ref{sec:closedloop}.

Given a bounded disturbance, the goal of the peak-to-peak control is to bound the reachable set (i.e., the closure of the set of all states reachable from the initial conditions given a bounded disturbance) by an invariant ellipsoid (defined below) and thereby bound the output.

\begin{definition}\label{def1}
The ellipsoid 
\begin{equation}
\label{eq:e_x}
    \varepsilon_x = \{x\in\mathbb R^n: x^T P^{-1}x\le 1\}, \quad P\succ 0,
\end{equation}
centered at the origin is (state) invariant for the dynamic system \eqref{eq:ss_open}, if the condition $x(0)\in\varepsilon_x$ implies $x(t)\in\varepsilon_x$ for all time instants $t\ge 0$.
\end{definition}

If $\varepsilon_x$ is the invariant ellipsoid define by matrix $P$, the output $z,\,\forall x(0)\in \varepsilon_x$ belongs to the ellipsoid define by $CPC^T$:
\begin{equation}
\label{eq:e_z}
    \varepsilon_z = \{z\in\mathbb R^m: z^T (CPC^T)^{-1}z\le 1\}.
\end{equation}

\subsubsection{Open-loop system}
\begin{figure}
\centering
\begin{subfigure}{.24\textwidth}
  \centering
  \includegraphics[width=1\linewidth]{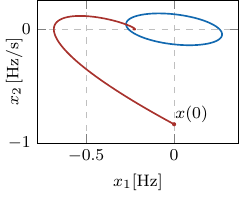}
  \caption{Without initial constraint}
  \label{fig:Elip1}
\end{subfigure}%
\begin{subfigure}{.24\textwidth}
  \centering
  \includegraphics[width=1\linewidth]{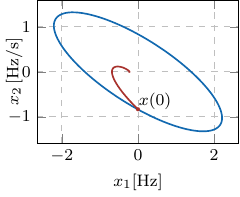}
  \caption{With initial constraint}
  \label{fig:Elip2}
\end{subfigure}
\caption{Bounding ellipse and state trajectory of the open-loop system.}
\label{fig:Elip}
\end{figure}

We first assess the minimum invariant ellipsoid of the open loop system, i.e., $K= 0$ in \eqref{eq:ss_open}, in terms of $\mathrm{tr}(CPC^T)$ where $\mathrm{tr}(\cdot)$ the trace of a matrix. Without generality, we considered the bounded disturbance $||\eta(t)||\le 1,\,\forall t\ge 0$, where $||\cdot||$ denotes the Euclidean norm. It can be easily maintained by scaling the system \eqref{eq:ss_general} with a factor of ${1}/{\Delta P_L}$. The matrix $P$ that leads to the minimum invariant ellipsoid can be obtained by evaluating the strictly convex function $\phi(\alpha) = \mathrm{tr}(CP(\alpha)C^T)$ over $0<\alpha<-2\max(\Re(\mathrm{spec(A)}))$, with $P$ satisfying the following Lyapunov equation \cite{p2p}:
\begin{equation}
\label{eq:Lyapn}
    AP + PA^T +\alpha P + \frac{1}{\alpha} EE^T = 0.
\end{equation}
The result of directly applying the above conclusion with $C$ being set as $\begin{bmatrix}
    1 & 0
\end{bmatrix}$ is showcased in Fig.~\ref{fig:Elip1}, where the red curve and blue ellipse represent the state trajectory and the minimum invariant ellipse respectively (the units of $x_1$ and $x_2$ axes are converted from per unit to SI). However, since the initial point is outside the ellipse, the state trajectory is not bounded by the invariant ellipse. To find the actual invariant ellipse, by which the state trajectory is bounded, the initial constraint $x(0)^TP^{-1}x(0)\le1$ has to be considered together with \eqref{eq:Lyapn}. This initial constraint can be further reduced to the following LMI using Schur lemma \cite{horn2012matrix}:
\begin{equation}
    \begin{bmatrix}
        1 & x(0)^T\\
        x(0) & P
    \end{bmatrix} \succeq 0.
\end{equation}
The result in this case is depicted in Fig.~\ref{fig:Elip2}. Although the state trajectory is bounded by the ellipse, the results become over-conservative as the maximum frequency deviation provided by the ellipse becomes much larger than that of the actual trajectory. Note that the two sub-figures have different scales for clarity and the two red curves are identical.

\begin{figure}
\centering
\begin{subfigure}{.24\textwidth}
  \centering
  \includegraphics[width=1\linewidth]{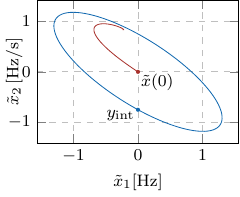}
  \caption{$a=1$}
  \label{fig:Elip3}
\end{subfigure}%
\begin{subfigure}{.249\textwidth}
  \centering
  \includegraphics[width=1\linewidth]{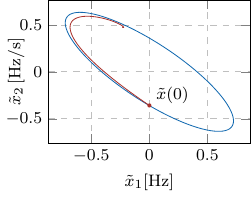}
  \caption{$a=a^*$}
  \label{fig:Elip4}
\end{subfigure}
\caption{Bounding ellipse and state trajectory after change of coordinates.}
\label{fig:Elip34}
\end{figure}

To reduce the conservativeness caused by the large initial condition $\dot \omega(0^+)$, system \eqref{eq:ss_general} is shifted to new coordinates:
\begin{subequations}
\label{eq:Tilde_x}
\begin{align}    
    \Tilde{x} &= x - x_0\\
    x_0 &=\underbrace{\begin{bmatrix}
        0 & \frac{a}{M_g}
    \end{bmatrix}^T}_{X} \Delta P_L,
\end{align}
\end{subequations}
where $a\in[0,\,1]$ is a scaling factor to achieve the least conservativeness. With the definition in \eqref{eq:Tilde_x} and $K = 0$ system \eqref{eq:ss-dx} can be rewritten in the form:
\begin{equation}
    \dot{\Tilde x} = A\Tilde{x}+(AX+E)\eta
\end{equation}
Clearly, $a = 0$ implies an unchanged coordinates (Fig.~\ref{fig:Elip1}), whereas $a = 1$ leads to a state trajectory with $\Tilde{x}(0)$ located at the origin (Fig.~\ref{fig:Elip3}). Although the second case gives a bounded trajectory, the result is still considerably conservative. Therefore, the following algorithm is proposed to determine the optimal $a$. Starting at $a= 1$, in each iteration gradually decrease $a$ until $\Tilde{x}_2(0,a)$ exceed $y_{\mathrm{int}}(a)$, and set the optimal scaling factor $a^*$ to be the value of $a$ in the previous iteration, where $\Tilde{x}_2(0,a)$ is the initial condition of $\Tilde{x}_2$ as a function of $a$ and $y_{\mathrm{int}}(a)$ is the intersection between the ellipse and the negative $\Tilde{x}_2$ axis as shown in Fig~\ref{fig:Elip3}. The result associated with $a = a^*$ is illustrated in Fig.~\ref{fig:Elip4}, where the conservativeness is significantly reduced with the state trajectory close to, yet bounded by the invariant ellipse.

\subsubsection{Closed-loop system} \label{sec:closedloop}
\begin{figure}[!b] 
	\centering
	\scalebox{0.9}{\includegraphics[]{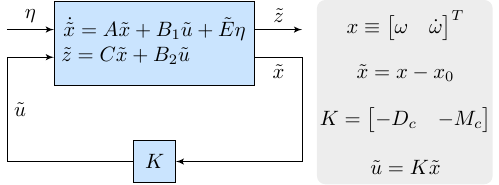}}
	\caption{State feedback control for adaptive VSM approach.}
	\label{fig:P2P_general}
	\vspace{-0.3cm}
\end{figure}
With the above strategy to determine $a^*$, we focus on the closed-loop system under the new coordinates. Note that due to the physical interpretation (virtual inertia and damping), the original control law $u = Kx$ defined in \eqref{eq:ss_g2} should be used, to ensure the system dynamics in the new coordinates remain unchanged as in \eqref{eq:ss_general}. Combining \eqref{eq:Tilde_x} and \eqref{eq:ss_open} gives the system dynamics with state $\Tilde{x}$:
\begin{subequations}
\begin{align}
    \dot{\Tilde x} & = A\Tilde{x}+B_1 \underbrace{K \Tilde{x}}_{\Tilde{u}}  +\underbrace{(AX+E+B_1KX)}_{\Tilde{E}}\eta  \\
    \Tilde{z} &= \underbrace{(C+B_2K)}_{C_{cl}}\Tilde{x},
\end{align}
\end{subequations}
where the two extra terms in $\Tilde{E}$ are due to the change of coordinates. The overall block diagram of the system is also depicted in Fig.~\ref{fig:P2P_general}, where the state feedback control is defined as $\Tilde{u} = K\Tilde{x}$.

\begin{figure}[!b]
    \centering
	\scalebox{1.2}{\includegraphics[]{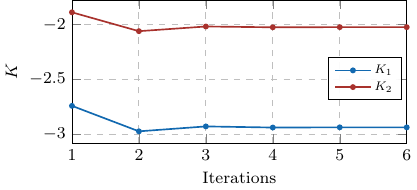}}
	\caption{Iteration progress of control gains in Algorithm~\ref{alg:1}.}
	\label{fig:K_plot}
	\vspace*{-0.3cm}
\end{figure}

The optimal control gain that minimizes $\mathrm{tr}(C_{cl}P{C_{cl}^T})$ can be determined by solving the following optimization problem~\cite{p2p}:
\begin{subequations}
\label{eq:SDP_dep}
    \begin{align}
        \min_{P,Y,Z} \quad & \mathrm{tr}(CPC^T + CY^TB_2^T + B_2YC^T + B_2 ZB_2^T)\\
        \mathrm{s.t.} \quad\; & \begin{bmatrix}
            AP+PA^T +\alpha P + B_1 Y + (B_1Y)^T & \Tilde{E}\\
            \Tilde{E}^T & -\alpha I
        \end{bmatrix}\preceq 0 \\
        & \begin{bmatrix}
            Z & Y\\
            Y^T & P
        \end{bmatrix}\succeq 0.
    \end{align}
\end{subequations}
and the optimal state feedback control gain is given by:
\begin{equation}
    \label{K_opt}
    K = YP^{-1}.
\end{equation}
However, due to the dependence of $A$, $B_1$ and $\Tilde{E}$ on $K$, \eqref{eq:SDP_dep} cannot be directly solved as an SDP as demonstrated in \cite{p2p}. To overcome this challenge, Algorithm~\ref{alg:1} is proposed, the main idea of which is summarized here. Initialize the system to the open loop [$K^{(0)} = 0$]. In each iteration, solve \eqref{eq:SDP_dep} with $K$ in  $A$, $B_1$ and $\Tilde{E}$ being viewed as a parameter [equaling the value obtained in the previous iteration, i.e., $K = K^{(k-1)}$]. The algorithm terminates if the error between two successive iterations becomes smaller than the pre-defined threshold $\epsilon$. However, the complexity of the combined SDP equations makes it difficult to derive an analytic convergence criterion for Algorithm~\ref{alg:1}. Fig.~\ref{fig:K_plot} illustrates the performance of Algorithm~\ref{alg:1} through the value of optimal control gains during iterations. It can be observed that the convergence is achieved within a few iterations, for the convergence error thresholds $\epsilon = 10^{-6}$. 

\begin{algorithm}[!t]
\caption{Iterative computation of optimal control gains}
\label{alg:1}
\begin{algorithmic}[1]
    \State Set $k=0$ and $\varepsilon^{(0)}=0$
    \State Initialize the system in open loop 
    \Comment $K^{(0)}=0$
    \While {$\varepsilon^{(k)}>\epsilon$ or $k=0$}
        \State $k = k + 1$
        \State Solve \eqref{eq:SDP_dep} with $K = K^{(k-1)}$ 
        \Comment derive $Y^{(k)},\,P^{(k)}$
        \State Compute $K^{(k)} = Y^{(k)}{P^{(k)}}^{-1}$
        \State Compute error
        \Comment $\varepsilon^{(k)}=\left|K^{(k)}-K^{(k-1)}\right|$
	\EndWhile
	\State Return $K^{(k)}$
\end{algorithmic}
\end{algorithm}

The optimal virtual inertia and damping from IBRs can therefore be determined according to \eqref{eq:ss_g2}:
\begin{subequations}
    \begin{align}
    \label{eq:DM_opt}
        \begin{bmatrix}
            D_c & M_c
        \end{bmatrix} = \begin{bmatrix}
            -K_1 & -K_2
        \end{bmatrix}.
    \end{align}
\end{subequations}
With the optimal control gain being determined, together with the bounded state invariant ellipsoid, the bound for the control input $\Tilde{u}$ can also be revealed \cite{p2p}:
\begin{equation}
\label{eq:u_max}
    ||\Tilde{u}|| = ||K\Tilde{x}|| \le \max_{\Tilde{x}P^{-1}\Tilde{x}\le 1} ||K\Tilde{x}|| = ||KP^{1/2}||.
\end{equation}
However, for the frequency control problem, one of the interests lies in identifying the maximum power injection of the IBRs during the entire frequency event, i.e., $t\in \mathcal{T}$, required by the control law, such that enough headroom can be reserved during the system scheduling process. Based on \eqref{eq:u_max} the bound of the original control input $\Delta P_c(t) = Kx(t) = K(\Tilde{x}(t)+x_0)$ in \eqref{eq:ss_open} can be derived as follow:
\begin{align}
\label{eq:max_P_c}
   \max_{t\in \mathcal{T}} |\Delta P_c(t)| & \le \max_{\Tilde{x}P^{-1}\Tilde{x}\le 1} |K\Tilde{x}+ Kx_0| \nonumber \\
    & = ||KP^{1/2}|| + |K x_0|.
\end{align}
The equality holds since in a practical system the maximum of $\Delta P_c$ always occurs between $t = 0$ and $t = t_m$, during which $K\Tilde{x}(t)$ and $Kx_0$ have the same sign. Similarly, the bound of frequency nadir can be developed as:
\begin{equation}
\label{eq:max_w}
    \max_{t\in \mathcal{T}} |\omega (t)|\le \max_{\Tilde{x}P^{-1}\Tilde{x}\le 1} \left| \underbrace{\begin{bmatrix}
        1 & 0
    \end{bmatrix}}_{e_1} \begin{bmatrix}
        \Tilde{x}_1 \\
        \Tilde{x}_2 
    \end{bmatrix}\right | = \sqrt{e_1Pe_1^T}.
\end{equation}
Equations \eqref{eq:max_P_c} and \eqref{eq:max_w} demonstrate the relationship between the required headroom of IBR (control effort) and the bound of the frequency nadir (performance), i.e.,  for a given power system, the optimal virtual inertia and damping determined in \eqref{eq:DM_opt} are able to bound the frequency nadir by \eqref{eq:max_w} with the required power reserve bounded by \eqref{eq:max_P_c}. 

\section{Frequency Constrained System Scheduling}\label{sec:3}
In order to utilize the control flexibility of IBR and achieve more efficient system stability support, this section aims to combine system scheduling and optimal control design, the decisions of which are conventionally made separately. Moreover, different from some of the existing works, e.g., \cite{9066910,shen2023optimal,chu2024scheduling} where the determination of the optimal IBR control parameters potentially makes the nonlinear frequency constraint even more complex, the proposed method, instead, \textcolor{black}{tries to identify the required headroom of IBRs during the optimization and the detailed control parameters can be revealed by solving a simple SDP after the optimization.}
This is achieved by extracting the relationship between the bounds of performance and control effort through the peak-to-peak control design developed in Section~\ref{sec:2}. 

\subsection{Problem Formulation}
The goal of the frequency-constrained system scheduling is to determine the optimal generator status and setpoints as well as the frequency control gains of IBRs for operation cost minimization while ensuring the system frequency constraints. We first write the concerned problem in the following general form:
\begin{subequations}
\label{eq:opt1}
\begin{align}
    \min_{y, K} & \quad f(y) \label{eq:cost} \\
    \mathrm{s.t.} & \quad g_1(y)\le 0, \quad g_2(y) =  0 \label{eq:g}\\
    & \quad h(y,K)\le 0 \label{eq:h}\\
    & \quad \dot x = A(y)x+B_1(y)u+E(y) \eta \label{eq:ss_opt}\\
    & \quad z = Cx \\
    & \quad u = Kx \label{eq:u_opt}\\
    & \quad \max_{t\in\mathcal{T}} |z_i(t)| \le \bar z_{i},\,\,\forall i ,\label{eq:z_max}
\end{align}
\end{subequations}
where \eqref{eq:cost} and \eqref{eq:g} represent the objective function and constraints in the conventional system scheduling problem (e.g., generator costs and operational constraints respectively) with $y$ being the conventional decision variables such as the generator status and power outputs (detailed expressions can be found in \cite{7833096}); $\eqref{eq:ss_opt}$-$\eqref{eq:u_opt}$ define the system frequency dynamics, frequency performance and control input respectively; \eqref{eq:z_max} restricts the peak value of different performance metrics among the frequency event ($t\in\mathcal{T}$), e.g., maximum RoCoF and maximum frequency deviation, with $\bar z_i$ being their maximum admissible values; \eqref{eq:h} is the constraint that confines the selection of the IBR frequency control gains $K$, given the operating condition $y$. An example of \eqref{eq:h} can be written as follows:
\begin{subequations}
\begin{align}
    \max_{t\in\mathcal{T}} |P_c + K_{1c} \omega(t) + K_{2c} \dot\omega(t)| & \le \bar P_{c},\,\forall c\in \mathcal{C} \label{eq:max_Pc} \\
    \sum_{c\in\mathcal{C}} K_{1c} &= K_1 \label{eq:sum_K1} \\
    \sum_{c\in\mathcal{C}} K_{2c} &= K_2, \label{eq:sum_K2}
\end{align}
\end{subequations}
where $P_c$ is the pre-disturbance power output of IBR $c$ [equivalent to $y$ in \eqref{eq:opt1}]; $K_{1c}$, $K_{2c}$ and $\bar P_c$ are their frequency control gains and the IBR power limit respectively. Constraint \eqref{eq:max_Pc} ensures the maximum power injection during the frequency event, including the steady-state and additional frequency support, from IBR $c$ does not exceed the power limit. The relationship between the individual control gains and the system ones is illustrated in \eqref{eq:sum_K1} and \eqref{eq:sum_K2}.

Note that though the conventional UC model is presented in this work as an example of the problem \eqref{eq:opt1}, it would also fit the market-based operation by considering the offers and bids.

\subsection{Existing Solutions}
The challenge of solving the problem \eqref{eq:opt1} lies in two aspects. First, the dependence of the system dynamics, i.e., $A$ and $B$ matrices, on the operating conditions ($y$) complicates the optimal control design significantly. Second, the expression of the closed-loop system performance and the control effort becomes extremely tedious and sometimes an explicit expression may be unavailable.

To overcome these challenges, two lines of work have been proposed in the literature. One is to discretize the dynamic system \cite{platbrood2013generic,golshani2019real,javadi2021frequency}, which enables significant simplification of constraints \eqref{eq:z_max} and \eqref{eq:h} at the cost of dramatically increased decision variables and constraints. The other one, on the contrary, does not need to include the system dynamic constraints \eqref{eq:ss_opt}-\eqref{eq:u_opt} in the optimization, by finding the mathematical expression of \eqref{eq:z_max} and \eqref{eq:h}. However, due to their highly nonlinear structure, these constraints require further reformulation to be incorporated into the optimization model, which may suffer from computational burden due to the introduction of binary variables or piecewise linearization. To further simplify the problem, the control parameters are typically assumed to be known and fixed in both approaches. 

Also, over-conservativeness can sometimes be inevitable. Taking \eqref{eq:max_Pc} as an example, finding the maximum of the IBR power injection (left-hand side) requires the substitution of the time instant when this term reaches maximum into the expression of frequency deviation and RoCoF, which becomes even more complicated than the frequency nadir constraint itself. In the existing research, this type of constraint is typically relaxed in the following way:
\begin{align}
\label{P_c_relax}
    \max_{t\in\mathcal{T}} |P_c + K_{1c} \omega(t) + K_{2c} \dot\omega(t)| &  \le \nonumber \\
    |P_c| + \max_{t\in\mathcal{T}} |K_{1c} \omega(t)| + \max_{t\in\mathcal{T}} |K_{2c} \dot\omega(t)| & \le \nonumber \\
    |P_c| + |K_{1c} \omega_{\mathrm{lim}}| + |K_{2c} \dot\omega_{\mathrm{lim}}| & \le \bar P_{c}.
\end{align}
The over-conservativeness comes from the facts that the maximum frequency deviation and RoCoF do not occur simultaneously (triangular inequality) and that the frequency nadir and maximum RoCoF constraints are typically not binding at the same time (replacing $\max_{t\in\mathcal{T}}(\cdot)$ with their limits). 

\subsection{Proposed Method} \label{sec:3.3}
In this section, we propose a method to solve \eqref{eq:opt1} by extracting the relationship between the bounds of performance and control effort obtained from the developed peak-to-peak control design in an offline manner. \textcolor{black}{Moreover, the determination of the IBR control parameters is converted to that of the IBR capacity reserve, thus well-fitting the existing framework of ancillary service provision.}

According to the derivation in \eqref{eq:max_P_c} and \eqref{eq:max_w}, the required headroom from IBRs ($\max_{t\in \mathcal{T}} |\Delta P_c(t)|$) is linked to the bound of the frequency metric ($\max_{t\in \mathcal{T}} |\omega (t)|$) through the matrix $P$. A different $P$ would result in a different relationship between these two quantities. This trade-off between the control effort and performance can be further leveraged by tuning the performance matrices $C$ and $B_2$. As a result, for a given operating condition ($y$), an IBR headroom can be determined with which the frequency metric can be maintained. With this method, the original decision associated with the optimal control gains is now changed to the optimal required headroom of IBRs, thus significantly reducing the complexity of constraints \eqref{eq:h} and \eqref{eq:z_max}. The optimization problem associated with the proposed method is given in the form below:
\begin{subequations}
\label{eq:opt2}
\begin{align}
    \min_{y, \bar u} & \quad f(y) \\
    \mathrm{s.t.} & \quad g(y)\le 0 \\
    & \quad l(y,\bar u)\le 0 \label{eq:l}\\
    & \quad \bar u \ge m(y) \label{eq:m},
\end{align}
\end{subequations}
where $\bar u$ is the bound of control effort and \eqref{eq:l} restricts the control effort according to the current operating conditions and the control limits, being equivalent to \eqref{eq:h} in the previous formulation. Since it is the IBR headroom ($\bar u$) instead of the control gains ($K$) that is determined in the optimization, \eqref{eq:l} can be written in the following form, which transforms the highly nonlinear constraint \eqref{eq:max_Pc} into the linear form:
\begin{subequations}
    \label{eq:max_Pc_linear}
\begin{align}
    |P_c + \bar u_c| & \le \bar P_c, \,\forall c\in\mathcal{C} \\
    \sum_{c\in\mathcal{C}} \bar u_c & = \bar u,
\end{align}
\end{subequations}
where the control gains and the system dynamic quantities associated with $t$ are replaced with $\bar u_c$, the steady-state capacity headroom of IBR $c$. Constraint \eqref{eq:m} ensures enough control capability is reserved for the system frequency constraints, where $m(y)$ is defined as the minimum IBR capacity headroom that would lead to a frequency nadir within the permissible range, in the operating condition $y$. The quantity $m$ as a function of $y$ is determined by evaluating the relationship between the bounds of control effort and frequency metric at different operating conditions, as demonstrated in the next subsection. 

\subsection{Peak-to-Peak Control as Operational Constraints} \label{sec:3.4}
\begin{figure}[!t] 
	\centering
	\scalebox{1.06}{\includegraphics[]{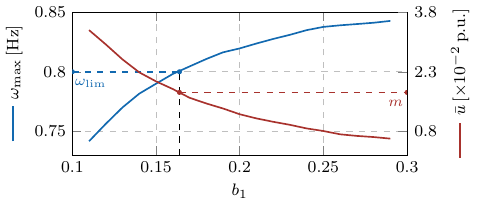}}
	\caption{Trade-off between the performance and control effort.}
	\label{fig:P2P}
\end{figure}

As illustrated in Section~\ref{sec:2}, for a given system dynamics and system performance, the peak-to-peak control design is able to determine the bounds of the control effort and the frequency metric under the optimal control gains. The trade-off between these two bounds can be leveraged by tuning the performance matrices $C$ and $B_2$. Without loss of generality, we set 
\begin{equation}
    C = \begin{bmatrix}
    1 & 0\\
    0 & 0
\end{bmatrix},\qquad B_2 = \begin{bmatrix}
    0\\
    b_1
\end{bmatrix}.
\end{equation}
The result of tuning $b_1$ is depicted in Fig.~\ref{fig:P2P}. It can be observed that as the weight of the control effort ($b_1$) increases, the optimal peak-to-peak control design tends to use less control input ($\bar u$), thus leading to an increased frequency nadir ($\omega_{\mathrm{max}}$). Therefore, the $b_1$, with which the frequency nadir reaches the limit ($\omega_{\mathrm{lim}}$) gives $m$ in this operating condition as defined in \eqref{eq:m}. As long as the actual IBR headroom is larger than or equal to $m$, the frequency nadir constraint can be maintained. 

\begin{figure}[!t] 
	\centering
	\scalebox{1.2}{\includegraphics[]{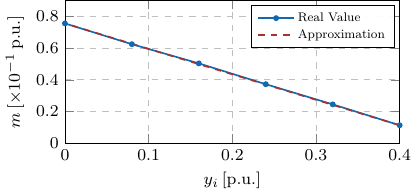}}
	\caption{Relationship between $m$ and $y$.}
	\label{fig:m_y}
\end{figure}

The above process is repeated for different operating conditions. As demonstrated in Section~\ref{subsec:2.2}, the frequency dynamics are influenced by the status of SGs. Hence, $m$ with various SG online capacities is evaluated. Since this work is targeted at the system with high IBR penetration, the number of SGs in the system is limited. Moreover, SGs with similar characteristics can be grouped together. An example of their impact on $m$ is showcased in Fig.~\ref{fig:m_y}, where $y_i$ represents the online capacity of SG group $i$. It is clear from the figure (blue curve) that less SG's online capacity requires more IBR's headroom to maintain the system frequency constraint. Furthermore, this curve can be approximated with a linear relationship accurately as indicated by the red dashed line. As a result, \eqref{eq:m} can be rewritten in the following form:
\begin{equation}
\label{eq:u_linear}
    \bar u \ge \sum_i k_i y_i + k_0,
\end{equation}
where $k_i$ and $k_0$ are the linear coefficients determined by linear regression. With \eqref{eq:u_linear} and \eqref{eq:max_Pc_linear}, the problem \eqref{eq:opt2} is not in linear form.

\section{Case Studies}\label{sec:4}
\begin{figure}[!b]
    \centering
    \vspace{-0.35cm}
	\scalebox{0.45}{\includegraphics[trim=0 0 0 0,clip]{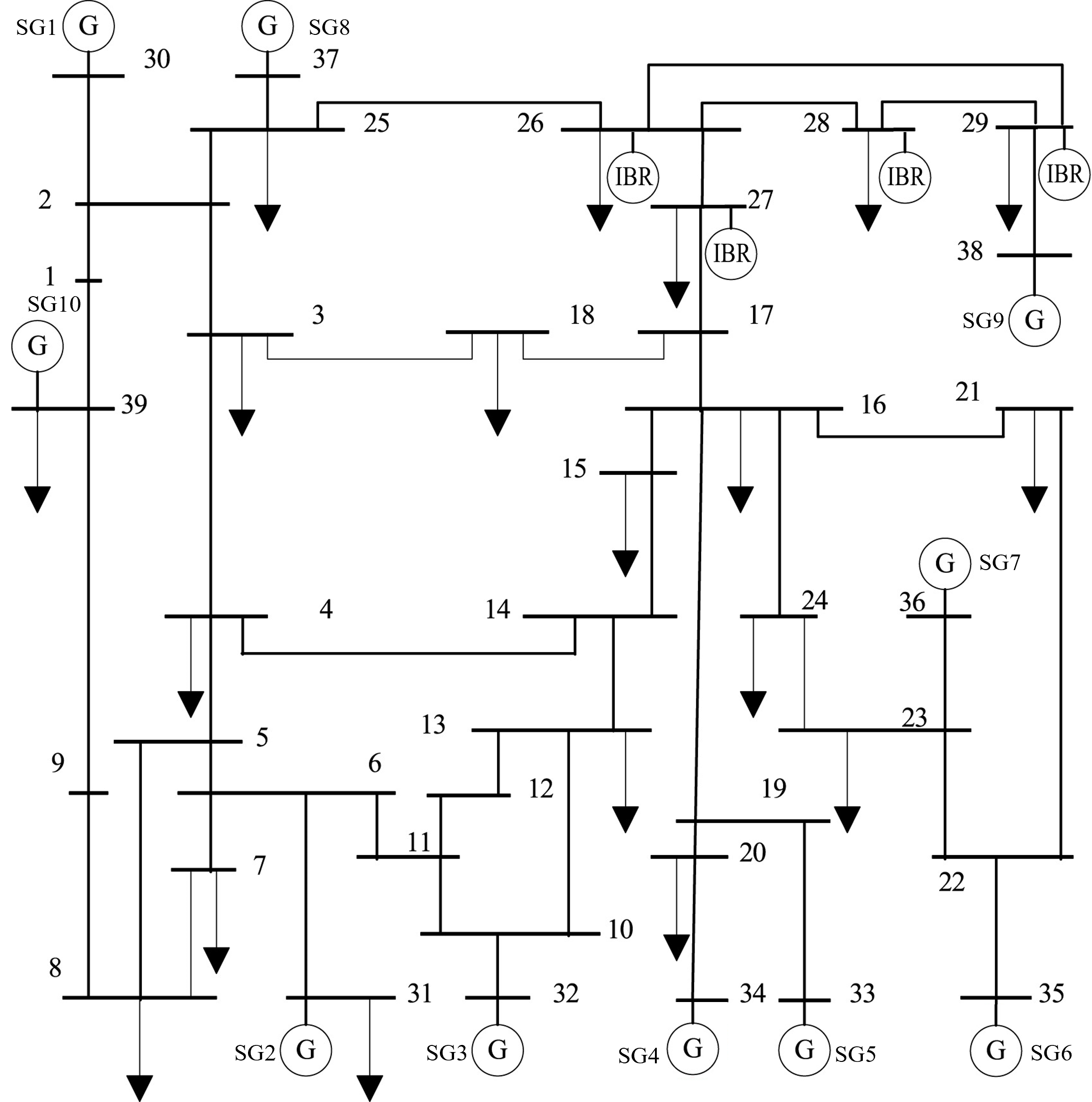}}
    \caption{\label{fig:39-bus}Modified IEEE-39 bus system.}
    \vspace{-0.35cm}
\end{figure}

To demonstrate the effectiveness of the proposed model, the modified IEEE-39 bus system shown in Fig.~\ref{fig:39-bus} is considered. IBRs with VSM-based frequency support capability are added at Bus 26, 27, 28 and 29 to increase the renewable penetration.
The parameters of transmission lines and loads are available in \cite{39_bus}. The load and renewable generation profile in \cite{9968474} is adapted for the simulation during the considered time horizon. 
Other system parameters are set as follows: load demand $P^D\in [5.16, 6.24]\,\mathrm{GW}$, base power $P_b = 8 \,\mathrm{GW}$ and maximum power loss $\Delta P_L = 800\,\mathrm{MW}$. The frequency limits set by National Grid are: $\Delta f_\mathrm{lim} = 0.8\,\mathrm{Hz}$, $\Delta f_\mathrm{lim}^\mathrm{ss} = 0.5\,\mathrm{Hz}$ and $\Delta \dot f_\mathrm{lim} = 1\,\mathrm{Hz/s}$. The MILP-based UC problem is solved by Gurobi (10.0.0) on a PC with Intel(R) Core(TM) i7-7820X CPU @ 3.60GHz and RAM of 64 GB.

\subsection{Performance of Peak-to-Peak Control Design}
\begin{figure}[!t]
    \centering
	\scalebox{1.2}{\includegraphics[trim=0 0 0 0,clip]{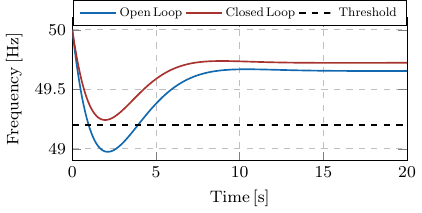}}
    \caption{\label{fig:f_plot}System frequency trajectories after a step disturbance.}
    \vspace{-0.35cm}
\end{figure}
The performance of the peak-to-peak control design applied to the frequency regulation in power systems with high IBR penetration is showcased in this subsection. A sample solution of the UC model is provided to the dynamical model resulting in the evolution of CoI frequency depicted in Fig.~\ref{fig:f_plot}.

The blue curve represents the frequency trajectory after the disturbance at $t = 0s$ in the open-loop system without any frequency support from IBRs. The frequency nadir in this case exceeds the Under-Frequency Load Shedding (UFLS) threshold (black dashed line) due to the reduced system inertia and frequency reserves from SGs, thus leading to frequency insecurity and severe economic losses for system operation. On the contrary, with the virtual inertia and damping provided by the IBRs, the closed-loop performance represented by the red curve significantly decreases the frequency nadir such that system frequency security can be ensured. The optimal feedback control gain determined by the peak-to-peak control design in this example is $K = [-2.08  -1.37]$.

\subsection{Comparison with Existing Formulation}
\begin{figure}[!b]
    \centering
    \vspace{-0.35cm}
	\scalebox{1.2}{\includegraphics[trim=0 0 0 0,clip]{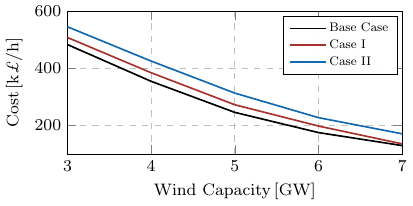}}
    \caption{\label{fig:Cost}Averaged system operation cost with different control methods.}
\end{figure}
The benefit of the proposed method where the frequency and the IBR reserve constraints are more efficiently incorporated into the system scheduling model in a simple linear manner is compared with the existing formulation, where the nonlinear frequency and IBR reserve constraints are dealt with via linear/second-order approximations, e.g., \cite{zhang2020modeling,chu2024scheduling}. The UC model covering 24 hours is carried out.

\subsubsection{\textbf{System operation cost}}
The averaged system operation cost is illustrated in Fig.~\ref{fig:Cost}. Different cases are defined as follows.
\begin{itemize}
    \item Base Case: without frequency constraints
    \item Case I: with frequency constraints, linearized using the proposed method
    \item Case II: with frequency constraints, linearized using analytical approximations
\end{itemize}
It is understandable that the Base Case presents the lowest operation cost among all cases since the frequency issue is not considered. Moreover, this cost gradually declines as the installed wind capacity in the system is increased. However, the frequency stability in this case cannot be guaranteed especially in high wind penetration, which may cause undesired events such as load shedding and even cascade failure. To ensure the system frequency stability, the frequency constraints are included in the optimization model while considering the virtual inertia and damping provision from IBRs. 

Applying the existing methods where the nonlinear constraints of frequency nadir and IBR headrooms are approximated analytically leads to the results represented by the blue curve (Case II). It can be observed that additional cost, varying from 30 to 70 $\mathrm{k\pounds}$ depending on the wind capacity, is inevitable to maintain the system frequency constraints. This cost increment mainly comes from the IBR reserve during normal operation so that virtual inertia and damping can be provided to the system when the contingency occurs. As a result, this additional cost decreases when there is more wind power in the system since the wind power otherwise being curtailed can now be utilized. In comparison, with the proposed method, the additional system operation cost to maintain the frequency constraints can be further reduced by more than half. At high wind penetration, the proposed method can even bring the cost down to the level close to the Base Case as more wind power can be utilized to provide frequency support. This is because the proposed method requires much less conservative IBR headroom compared with the existing method to achieve similar effectiveness in terms of system frequency support.

\subsubsection{\textbf{IBR headrooms}}
\begin{figure}[!t]
    \centering
	\scalebox{1.2}{\includegraphics[trim=0 0 0 0,clip]{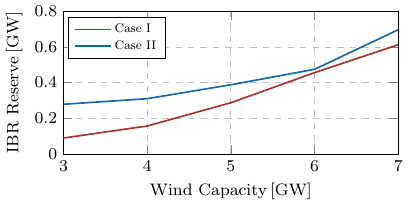}}
    \caption{\label{fig:IBR_reserve}IBR reserve for frequency regulation with different control methods.}
    \vspace{-0.35cm}
\end{figure}

The hourly-averaged IBR headrooms in Cases I and II are shown in Fig.~\ref{fig:IBR_reserve}. It can be observed that in both cases, a larger IBR reserve is required to maintain the system frequency stability when the wind capacity increases since less inertia and frequency response from SGs are available in the system. In addition, our proposed method represented by the red curve utilizes much less IBR reserve, in comparison with Case II at all wind penetration levels, which also justifies the cost saving illustrated in Fig.~\ref{fig:Cost}. However, it should be noted that the difference between the IBR reserve in the two cases does not directly link to the difference in the system operation cost in Fig.~\ref{fig:Cost}. For instance, the IBR headrooms in Case II increases with a faster trend compared with Case I before the wind capacity reaches $6\,\mathrm{GW}$ in Fig.~\ref{fig:IBR_reserve}, whereas the cost difference between the two cases in Fig.~\ref{fig:Cost} are almost the same. This is because, with the proposed method, more wind power can be utilized, which may reduce the number of online SGs, thus requiring more IBR reserve to maintain the frequency stability. Nevertheless, the overall system operation cost is always lower than Case I. 

\subsubsection{\textbf{Instantaneous power injection}}
\begin{figure}[!t]
    \centering
	\scalebox{1.2}{\includegraphics[trim=0 0 0 0,clip]{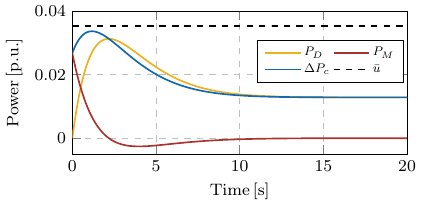}}
    \caption{\label{fig:p_plot}Power injection from IBRs to the grid.}
    \vspace{-0.35cm}
\end{figure}

One of the key challenges in the frequency-constrained optimization with the frequency support from IBRs is the identification of the feasible range of the frequency support parameter, e.g., the virtual inertia and damping. This is achieved by confining the total power injection from IBR to the grid. Therefore, to further demonstrate the utilization of the IBR reserve, the trajectories of different power injections from the IBR to the grid of a sample solution during the 24-hour scheduling are illustrated in Fig.~\ref{fig:p_plot}, where the powers are defined as:
\begin{equation}
    \Delta P_c = \underbrace{-D_c\cdot\omega}_{P_D} + \underbrace{(-M_c\cdot\dot\omega)}_{P_M}\,,
\end{equation}
and $\bar u$ is defined as in \eqref{eq:opt2}. Since $P_D$ and $P_M$ are proportional to the frequency deviation and RoCoF, their maximum values are attained at $t = t_m$ and $t = 0$ respectively. Depending on the trends of $P_D$ and $P_M$, the maximum of the total power injection ($\Delta P_c$) would occur in $ t\in [0,t_m]$, as indicated by the blue curve. In this case, since the initial increasing trend of $P_D$ dominates the decreasing trend of $P_M$ and vice versa latter, $\Delta P_c$ increases first and then decreases. Nevertheless, the maximum value of $\Delta P_c$ ($0.034\,\mathrm{p.u.}$) is only slightly higher than that of $P_M$ ($0.027\,\mathrm{p.u.}$) and $P_D$ ($0.031\,\mathrm{p.u.}$). The boundary of the control effort computed by the proposed method, \eqref{eq:max_P_c} is ($0.035\,\mathrm{p.u.}$), which ensures the conservativeness yet remains very close to the actual maximum, indicating a good approximation. On the contrary, the limit of the control effort computed according to the existing approach \eqref{P_c_relax}, in this case, is $0.061\,\mathrm{p.u.}$ (increased by about $75\%$), thus being over-conservative and leading to significant system operation cost.

\subsubsection{\textbf{Computational time}}
The computational time of the proposed method where the frequency nadir and IBR reserve constraints are replaced with a simple linear one and hence formulated as MILP is compared with the Mixed-Integer Second Order Cone Programming (MISOCP) in \cite{chu2024scheduling} with the results shown in Table~I. Note that the case without frequency constraints corresponds to the conventional UC problem, thus belonging to MILP. It is clear that the proposed method barely increases the computational time compared with the case without the frequency constraints since the frequency constraints are linear. It also significantly reduces the computational time due to the incorporation of the frequency constraints in the SOC formulation, demonstrating the effectiveness of the proposed model. 

\begin{table}[!b]
\renewcommand{\arraystretch}{1.2}
\caption{\textcolor{black}{Computational time of different formulations}}
\label{tab:time}
\noindent
\centering
    \begin{minipage}{\linewidth} 
        \renewcommand\footnoterule{\vspace*{-5pt}} 
        \begin{center}
            \begin{tabular}{ c || c | c  }
                \toprule
                  & \multicolumn{2}{c}{$\mathbf{Time\,[s]}$} \\
                \cline{2-3}
                & MILP & MISOCP \\ 
                \cline{1-3}
                \textbf{without frequency constraints} & 5.03 & N/A  \\
                \cline{1-3}
                \textbf{with frequency constraints} & 5.19 & 21.56  \\
               \bottomrule
            \end{tabular}
        \end{center}
        \end{minipage}
\end{table}

\subsection{Alignment with the conventional UC and ED framework}
In the conventional power system operation framework, UC problem involves determining the generator status over a certain period to meet the expected demand at the lowest possible cost. Once the units are committed, ED determines the optimal power output for each committed unit to meet the current demand after the uncertainty of the renewables and demand has been realized while adhering to operational constraints. The proposed method which converts the determination of virtual inertia and damping to the determination of the IBR capacity headrooms is well aligned with the conventional UC and ED framework. During the UC stage, only the IBR capacity reserve is determined, whereas the specific value of virtual inertia and damping required by the system can be determined at the ED stage when the uncertainty of the renewables and demand has been realized, thus utilizing the IBR control flexibility in a more efficient manner. 

\begin{figure}[!t]
    \centering
	\scalebox{1.2}{\includegraphics[trim=0 0 0 0,clip]{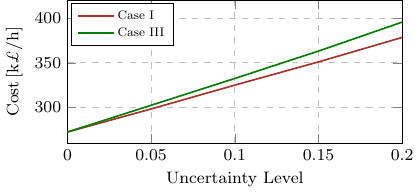}}
    \caption{\label{fig:uncertainty}System operation cost under different uncertainty level.}
    \vspace{-0.3cm}
\end{figure}

To demonstrate the benefit of uncertainty management, the proposed method (Case I) where the IBR reserve is determined in the first stage and the specific control parameters (virtual inertia and damping) are determined in the second stage is compared with the case where the IBR control parameters are determined in the first stage and remain the same in the second stage (Case III). Since the uncertainty modeling is not the focus of this work, we implement the  deterministic UC and explicit uncertainty level of the wind generation, represented by the the ratio of the standard deviation to the mean. The results are depicted in Fig.~\ref{fig:uncertainty}. To deal with the uncertainty, the system has to operate in more conservative conditions, thus leading to higher generation cost in both cases. However, with the proposed method (red curve), if there is more frequency response in the system at the ED stage due to the uncertainty, part of the IBR reserve can be released by using less virtual inertia and damping to produce energy, thus reducing the system operation cost. In addition, this trend becomes more obvious as the uncertainty level increases. 

\section{Conclusion} \label{sec:5}
This paper proposes a novel frequency-constrained system scheduling model, which replaces conventional highly nonlinear frequency nadir and IBR reserve constraints with a linear one by incorporating the peak-to-peak control design. During the scheduling stage, the IBR capacity reserve required by the optimal virtual inertia and damping is determined and can be defined as a grid service, whereas the exact control parameters can be revealed afterwards when closer to real time, benefiting from the software-defined nature of IBRs. This leads to less IBR headroom requirement, better uncertainty management, lower system operational cost, and faster computational time, demonstrated through detailed case studies.

\bibliographystyle{IEEEtran}
\bibliography{bibliography}
\end{document}